\documentclass[10pt]{article}

\usepackage{amsmath}
\usepackage{amssymb}

\usepackage{graphicx}

\usepackage{cite}

\usepackage{color} 

\usepackage[labelfont=bf,labelsep=period,justification=raggedright]{caption}

\makeatletter
\renewcommand{\@biblabel}[1]{\quad#1.}
\makeatother

\date{}

\begin{document}

\begin{flushleft}
{\Large
\textbf{The Interrupted Power Law and the Size of Shadow Banking}
}
\\
Davide Fiaschi$^{1}$, 
Imre Kondor$^{2}$, 
Matteo Marsili$^{3,\ast}$,
Valerio Volpati $^{4}$
\\
\bf{1} Dipartimento di Economia e Management, University of Pisa, Pisa, Italy
\\
\bf{2} Parmenides Foundation, Pullach b. M\"unchen, Germany
\\
\bf{3} The Abdus Salam International 
Centre for Theoretical Physics, Trieste, Italy
\\
\bf{4} International School for Advanced Studies (SISSA), 
Trieste, Italy
\\
$\ast$ E-mail: Corresponding author marsili@ictp.it
\end{flushleft}

\section*{Abstract}
Using public data (Forbes Global 2000) we show that the asset sizes for the largest global firms follow a Pareto distribution in an intermediate range, that is ``interrupted'' by a sharp cut-off in its upper tail, where it is totally dominated by financial firms. 
This flattening of the distribution contrasts with a large body of empirical literature which finds a Pareto distribution for firm sizes both across countries and over time. 
Pareto distributions are generally traced back to a mechanism of proportional random growth, based on a regime of constant returns to scale. This makes our findings of an ``interrupted'' Pareto distribution all the more puzzling, because we provide evidence that financial firms in our sample should operate in such a regime. 

We claim that the missing mass from the upper tail of the asset size distribution is a consequence of shadow banking activity and that it provides an (upper) estimate of the size of the shadow banking system. This estimate -- which we propose as a shadow banking index -- compares well with estimates of the Financial Stability Board until 2009, but it shows a sharper rise in shadow banking activity after 2010. Finally, we propose a proportional random growth model that reproduces the observed distribution, thereby providing a quantitative estimate of the intensity of shadow banking activity.

\section*{Introduction}

If we take the Forbes Global 2000 list as a snapshot of the global economy (see Materials and Methods), we find that financial firms dominate the top tail of the distribution of firms by asset size: the highest placed firm classified as non-financial is General Electric, which ranks 44th in the 2013 Forbes Global 2000 (FG2000) list. General Electric is also the highest placed non-financial firm in the 2013 Fortune 500 list, which covers only the US economy, where it ranks 11th. 
This seems to be a recent trend: General Electric was  the largest non-financial firm by asset size also in the 2004 FG2000 list and in the Fortune 500 list of 1995, but then it ranked 22nd and 3rd, respectively. 
Firm size can also be measured by other variables such as total sales, number of employees, or market value. However, these can be strongly affected by the fluctuations in market prices, and by the conditions of labor and financial markets; this is why we consider assets value a sounder proxy for the firm size.
Financial firms form approximately 30\% of the firms in the FG2000 list, and account for approximately 30\% of the total sales, profits and market value, a share that has been roughly constant in the whole period 2003-2012 studied. 
Yet, financial firms account for 70\% of total assets in the 2004 FG2000 list, a share that rose to 87\% in the 2013 list.

\begin{figure}[h]
   \centering
  \includegraphics[width=4.5in]{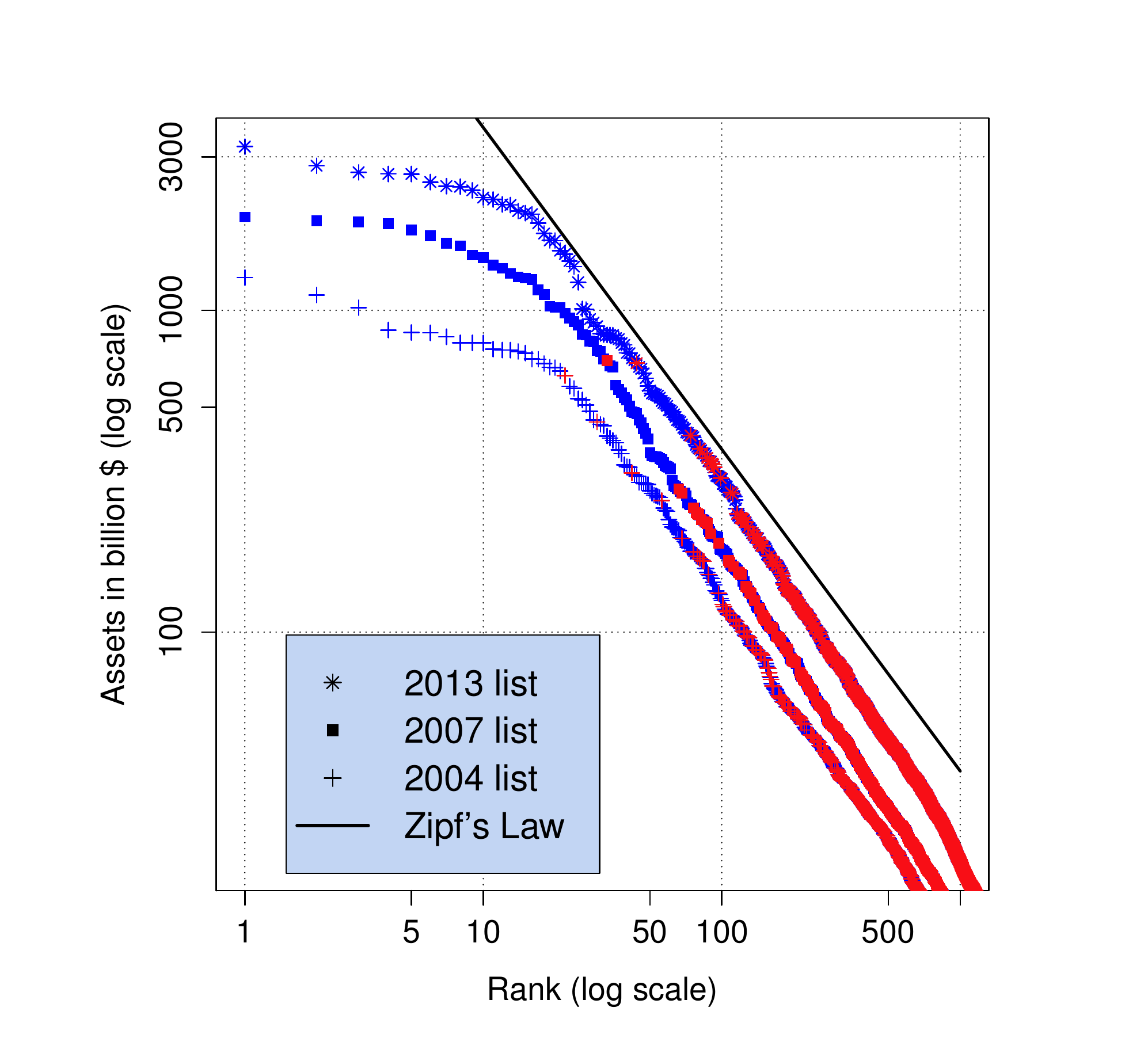} 
   \caption{Rank plot of the 2004 list ($+$), 2007 list ($\Box$) and 2013 list ($\ast$) of FG2000 by asset size. Financial firms are shown in blue, while the other firms in red. The straight line corresponds to Zipf's law and is drawn for comparison. }
   \label{fig:distr_asset_size_all}
\end{figure}

Besides being remarkable in themselves, the sizes of the biggest financial firms also display a peculiar distribution:
the 12th largest firm in the 2013 FG2000 list is the Royal Bank of Scotland with \$2.13 trillion in assets, which is comparable to the UK's 
gross domestic product (\$2.4 trillion). Yet its size is not much smaller than the largest firm in the list, Fannie Mae, which has assets worth  \$3.2 trillion. 
 This observation contrasts with the common view in the literature documented across countries and over time (see \cite{Axtell2001,Fujiwara2004,Gabaix2009}) that firm sizes $S$ follow a Paretian distribution as

\begin{equation}
\label{eq:power_law}
{\rm Prob}\{S\ge x\}\simeq c x^{-\gamma},
\end{equation}
with  $\gamma, c>0$.

Fig. \ref{fig:distr_asset_size_all} shows that the rank plot of the firms included in the 2004, 2007 and 2013 lists of FG2000 follows Eq. (\ref{eq:power_law}), with an exponent $\gamma$ close to one, corresponding to Zipf's law \cite{Axtell2001}, only from the 20th largest company downward. The upper tail, which is entirely dominated by financial firms, levels off. If Zipf's law were to hold also for the top 20 companies, we would expect Fannie Mae to be ten times as large as the Royal Bank of Scotland (\$21.3 instead of  \$3.2 trillion).

This anomaly in the shape of the top tail of the assets distribution is the starting point of our analysis.

From a theoretical point of view, the occurrence of power laws (i.e. Pareto distributions) in the size distribution of firms has been related to proportional random growth (PRG) models  \cite{BouchaudMezard2000,Gabaix2009,Sornette2013}. In what follows, we shall first enquires whether departures from the PRG model's prediction may be due to an anomalous dynamic of financial firms that dominate the upper tail of the distribution. 
We conclude that the available data suggest that PRG \textit{should} hold for financial firms.
The analysis therefore provides a theoretical framework which allows us to calculate the hypothetical  distribution of assets in the absence of any anomaly. Next, we argue that the difference between this hypothetical distribution and the actual one can be taken as a proxy for the size of the so-called \textit{shadow banking system}, which has been broadly defined as credit intermediation involving entities and activities outside the regular banking system (see \cite{FSB2012}, p. 3), and is the subject of much debate at the time of writing \cite{NewsEU2013,FSB2013}. Finally we discuss a simple generalization of the model proposed in Ref. \cite{Sornette2013}, which allows a first investigation of the determinants of the observed anomaly.

\begin{figure}[h]
\includegraphics[width=7cm]{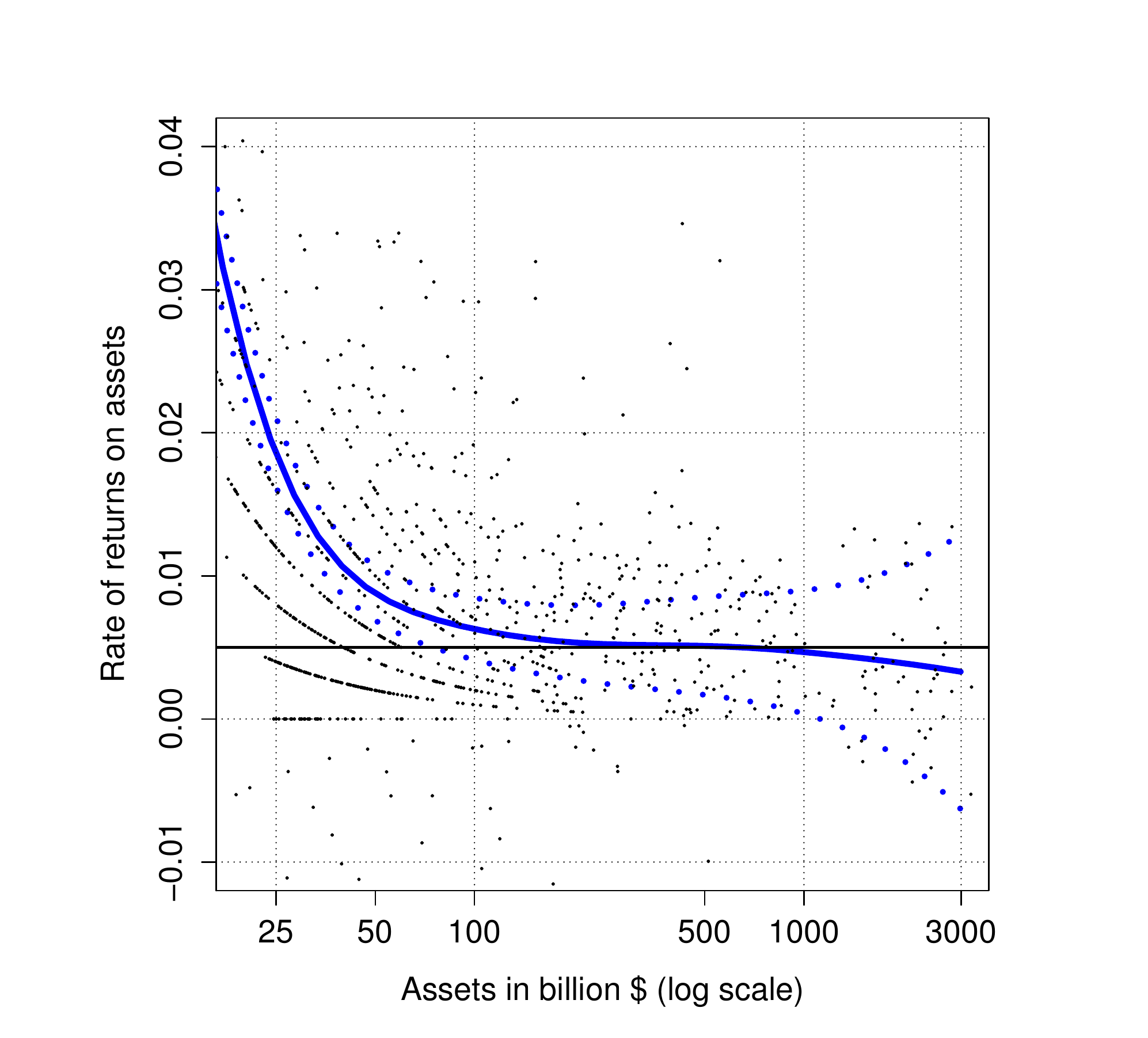}%
	\includegraphics[width=7cm]{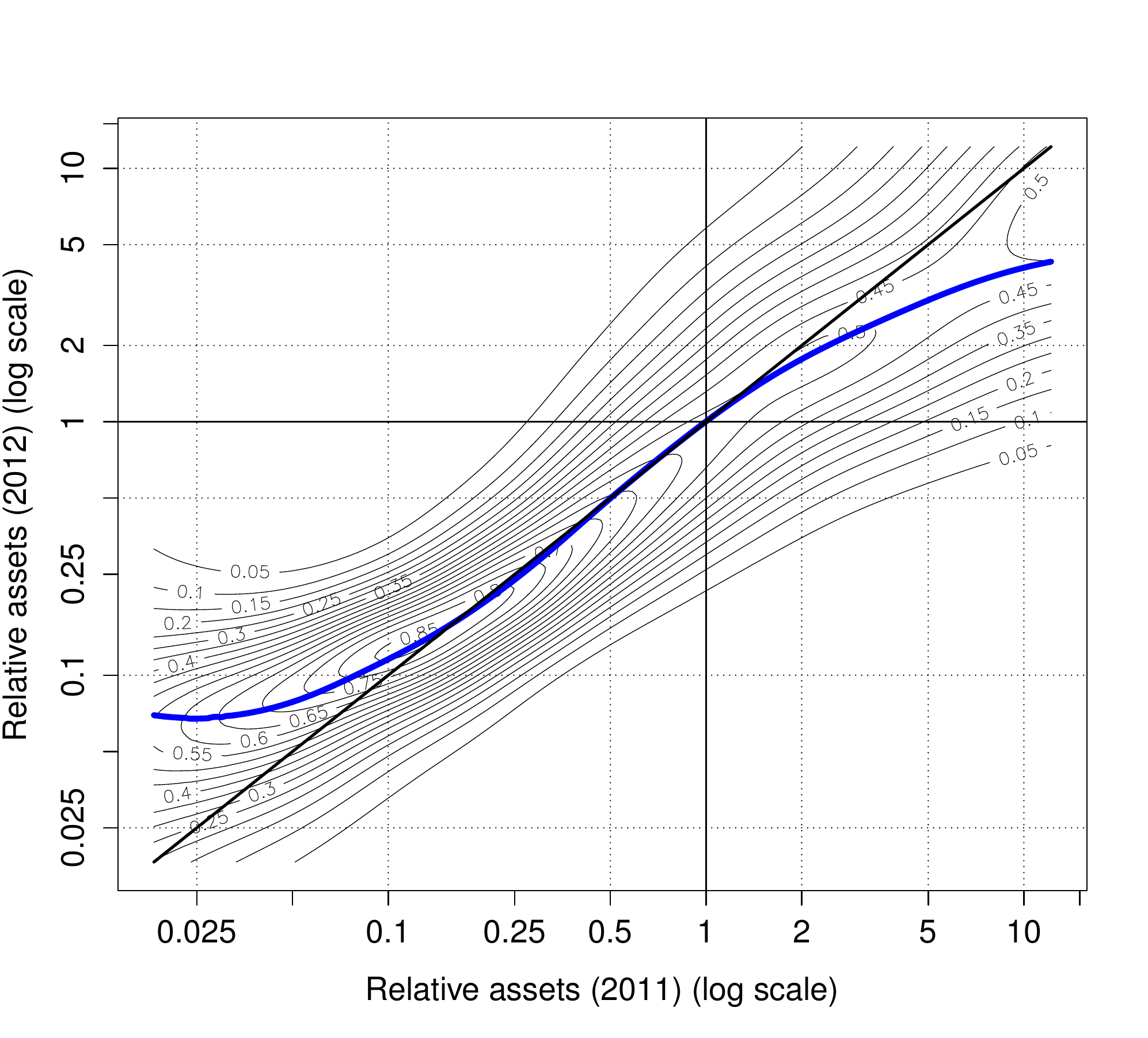}%
\caption{Left panel: non-parametric estimate of the relationship between the rates of return on assets and the levels of assets (bold line, dotted lines refer to $5\%$ confidence interval). Right panel: estimate of the stochastic kernel (i.e. of the conditional probability distribution), and of the expected level of assets in 2012 conditioned on the level of assets in 2011 (light grey lines and bold blue line respectively). Both plots refer to financial firms in the 2012 and 2013 FG2000 lists.}
\label{fig:retuns_to_scale_fin}
\end{figure}

\section*{Results}

\subsubsection*{Proportional Random Growth Model\label{sec:ProportionalRandom ModelforFinancialFirms}}

The observed Paretian distribution has generally been related to a mechanism of PRG which assumes that firms grow proportionally to their size (see \cite{Gabaix2009} p. 259, for more details).
In particular, a key empirical testable hypothesis of PRG models is that the rate of return on assets (i.e. the ratio of total profits to total assets) is independent of the level of assets, as it should be for industries with constant returns to scale. Firms in the financial sector are indeed expected to obey constant returns to scale. 
Our analysis of the FG2000 sample corroborates this hypothesis: 
the left panel in Figure \ref{fig:retuns_to_scale_fin} provides evidence for the flat relationship between the rates of return on assets for the years 2011 and 2012 (corresponding to the 2012 and 2013 lists of FG2000) and the level of total assets for most of the range of the assets distribution. 
In spite of this behavior of the rates of return on assets, the expected level of (relative) assets of firms at period $t$ conditioned on the level of (relative) assets at period $t-1$, is proportional to the latter only in an intermediate range. As shown in 
the right panel of Figure \ref{fig:retuns_to_scale_fin}, while banks of intermediate size grow proportionally to their size, the largest ones grow less than linearly (see caption of the figure). These findings are consistent with earlier results in \cite{WheelockWilson2012} and \cite{RestrepoEtAl2013}. 

\begin{figure}
\includegraphics[width=7cm]{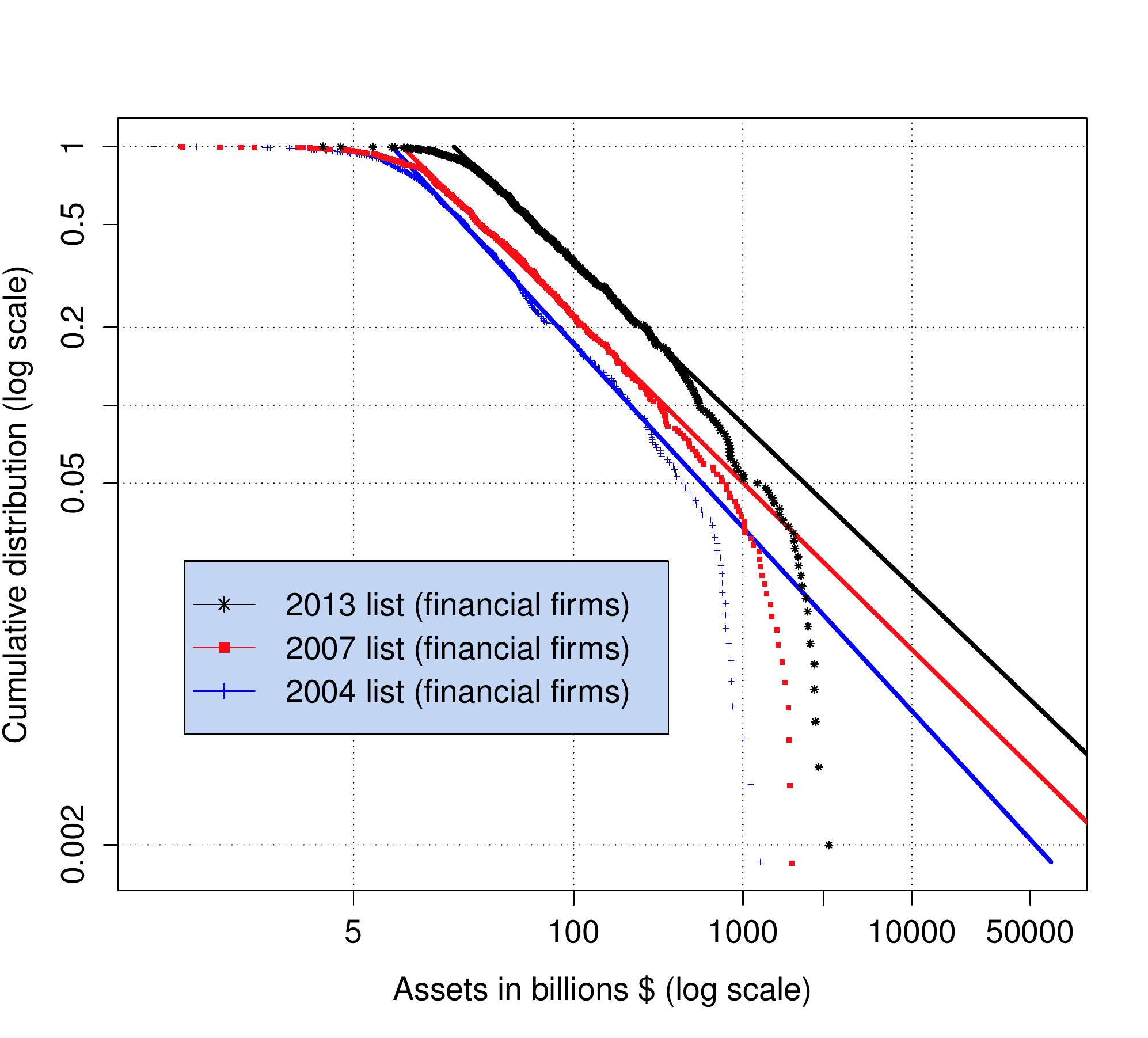}%
\includegraphics[width=7cm]{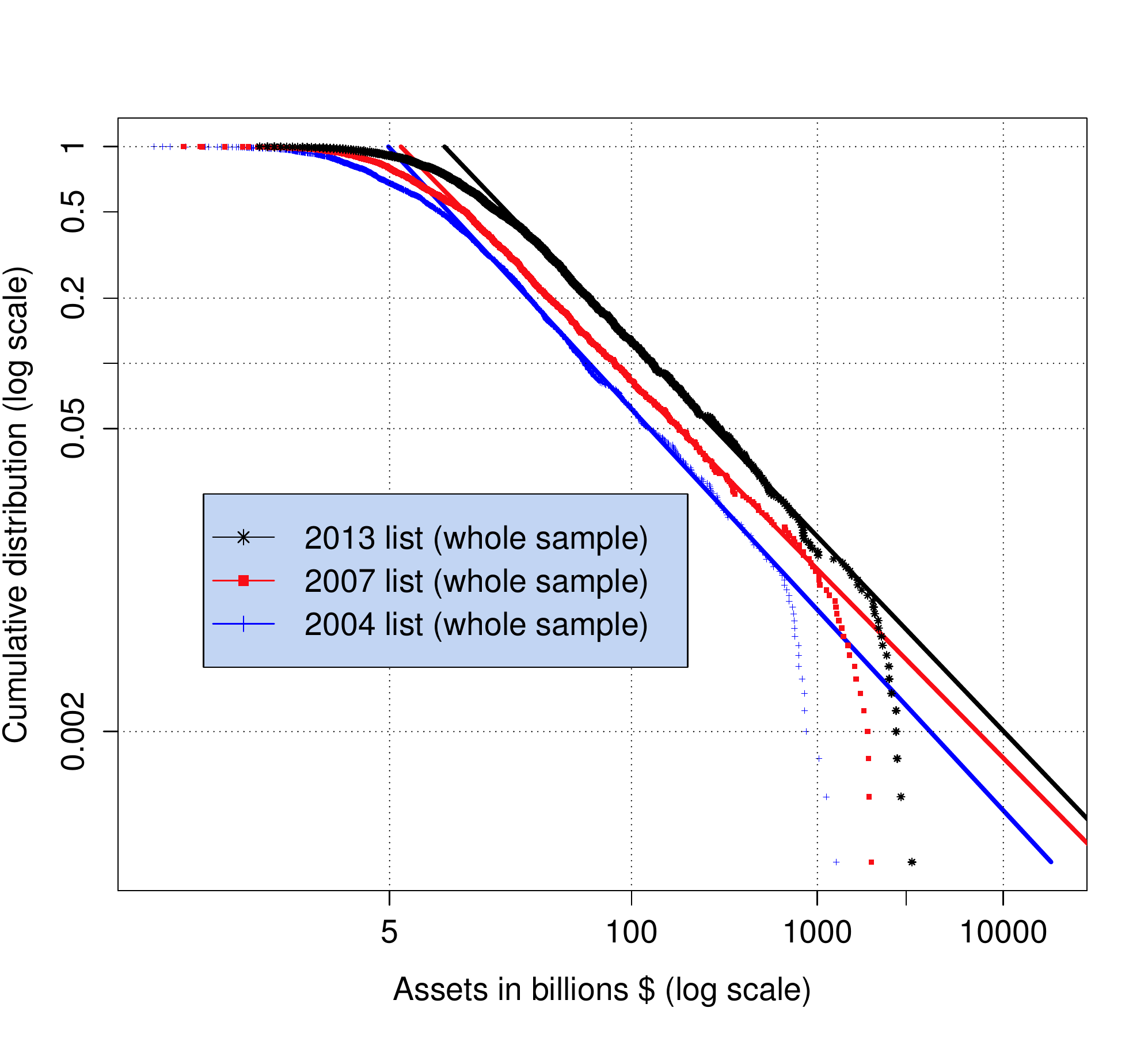}%
\caption{Cumulative distribution ${\rm Prob}\{S\ge x\}$ of asset sizes $S$  for financial (left panel) and all (right panel) firms in 2003, 2006, and 2012 (2004, 2006, and 2013 of FG2000 lists). The straight line is obtained as a linear fit in an intermediate range of $\log {\rm Prob}\{S\ge x\}$ vs $\log x$ (see Table \ref{tab:estimatedParetoCoeff}).}
\label{fig:estimateTheoreticalPowerLaw2003_2006_2012}
\end{figure}

The estimate of the rates of return shows no evidence of decreasing returns to scale for financial firms, thereby lending support to the PRG mechanism. However, the bending in the estimated expected level of assets highlights how the PRG inexplicably does not hold for the largest financial firms (more or less the top 13\% in the 2013 list). This finding is reflected in the distribution of asset sizes of financial firms, reported in the left panel of Fig. \ref{fig:estimateTheoreticalPowerLaw2003_2006_2012}, which follows a power law distribution in an intermediate range, but consistently bends downwards in the top tail. 
Such deviation from a theoretical power law behavior is much sharper than that occurring in the distribution of all  firms
(reported in the right panel of Fig. \ref{fig:estimateTheoreticalPowerLaw2003_2006_2012}). 

\cite{Cristelli2012} observed that Zipf's law (as with power laws in general) holds as a property of a system as a whole, but it may not hold for its parts. As such, it is manifest in samples that preserve a form of {\em coherence} (with the whole system), but fails to hold in incomplete samples that account for only part of the system (see \cite{Cristelli2012}).  Our findings of deviations from a power law behavior for financial firms which are more pronounced than for the whole economy, indicate that the Pareto distribution of asset sizes should be considered as a property that applies to the whole economy, rather than to a particular sector. This is consistent with empirical findings e.g. in  \cite{Axtell2001}, and suggests that, in the absence of anomalies, one should expect a hypothetical assets distribution that would perfectly obey the PRG predictions up to the largest firms.

Table \ref{tab:estimatedParetoCoeff} reports the ranges considered in the estimate of the power law distribution, and the estimate of the Pareto exponent $\gamma$ of Eq. (\ref{eq:power_law}) for all firms in the FG2000 list from 2004 to 2013 (2005 is missing for lack of data).
The estimated Pareto exponent for the whole sample $\hat{\gamma}$ peaks at the beginning of the period and steadily decreases until it reaches the lowest level in 2007 (2008 list), before the financial crisis. 
Then it increases suddenly in 2008 and remains relatively stable thereafter. 
Table \ref{tab:estimatedParetoCoeff} also reports the estimate of the 
Pareto exponent 
of the distribution of financial firms $\hat{\gamma}_{\rm fin}$ : $\hat{\gamma}_{\rm fin}$ is smaller than $\hat\gamma$ but it exhibits a behavior similar to  $\hat{\gamma}$, with the important exception that it starts to decline again after the crisis.

The estimated exponents $\hat{\gamma}$ and $\hat{\gamma}_{\rm fin}$ are both less than one for the whole period. The simplest PRG model predicts a Pareto exponent larger than one \cite{Gabaix2009}; however Bouchaud and Mezard \cite{BouchaudMezard2000} argue that $\gamma<1$ can be obtained within models of PRG with random shocks and trading of assets among firms if this trading is restricted in size and happens within a sparse network. Malevergne {\em et al.} \cite{Sornette2013} provide a different mechanism of PRG which explicitly accounts for the entry and exit of firms. In their paper an exponent smaller than one characterizes an economy where 
the accumulated resources of the economy are not channeled to investment in new enterprises but rather reinvested in existing firms \cite{Sornette2013}.

\subsubsection*{The Shadow Banking Index}
\label{sec:ShadowBankingIndex}

Shadow banking (SB) is a relatively new concept; the term itself is attributed to Paul McCulley \cite{McCulley2007}.  SB is a part of the wholesale money market where, in contrast to the regular banking system, it is not the central bank, but, at least in theory, private institutions that provide a backstop when necessary. This explains why SB has remained outside regulation (see, however, \cite{Fein2013}). During the 2007-08 crisis, which is often described as a run on the SB system \cite{PozsarEtAl2013}, this private guarantee proved insufficient, and without massive public intervention the collapse of the SB system would have brought down the whole global financial system. The first taxonomy of the different institutions and activities of SB was given by Pozsar \cite{Pozsar2008}, who also constructed a map to describe the flow of assets and funding within the system. The rise of a large part of SB was motivated by regulatory and tax arbitrage, and as such represented the answer of the finance industry to regulation, in particular to capital requirements. Other components responded to a real economic demand for different types of financial intermediation \cite{Mehrlingetal2013}.  Irrespective of the shortcomings or merits of the system, it is still true that shadow banking has remained by and large unregulated, its systemic risks implications uncharted, and its connections with the rest of financial system opaque. Indeed SB is, at the time of writing, one of the most important issues on the agenda of financial reform \cite{NewsEU2013,FSB2013}.

For us, the only property of interest of the SB system is its total volume. Estimates of its size differ in nature: Gravelle and Lavoie \cite{Gravelle2013} distinguish between two broad approaches to measuring the SB sector, one which is based on identifying the entities that contribute to it, and the other based on mapping the activities that constitute it. They also differ quantitatively, because of the difficulty to determine precisely which financial activities should be included in the calculation. For example, the Deloitte Shadow Banking Index \cite{Deloitte} shows a rise of the SB system in the US before 2008, but then displays a dramatic drop, suggesting that the phenomenon is now over. The index is built from specific components which are known to have played a major role in the crisis, and its decline after 2008 reflects the deflation of these markets. The Financial Stability Board (FSB) estimates that SB "[...] grew rapidly before the crisis, rising from \$26 trillion in 2002 to \$62 trillion in 2007. The size of the total system declined slightly in 2008 but increased subsequently to reach \$67 trillion in 2011" \cite{FSB2012}. 

Below we propose an index for the size of the SB system, denoted by $I_{SB}$, based on the idea that, in an ideal economy where finance operates in a regime of constant return to scale, the power law distribution should extend all the way to the largest firms. Since the top tail of the distribution is dominated by financial firms, we are led to attribute the mass missing from the distribution of asset sizes to "[...] credit intermediation involving entities and activities outside the regular banking system" \cite{FSB2012}, i.e. to shadow banking.
Fitting the middle range of the distribution to a power law behavior (as in the left panel of Fig. \ref{fig:estimateTheoreticalPowerLaw2003_2006_2012}) leads us to a theoretical estimate $\hat S_{[k]}$ of what the size of the $k^{\rm th}$ largest firm should be. Summing the difference between this theoretical estimate and the actual size $S_{[k]}$ of the $k^{\rm th}$ largest firm, over $k$, i.e.:
\begin{equation}
I_{SB}=\sum_{k=1}^N \left(\hat S_{[k]}-S_{[k]}\right)
\label{eq:indexSB}
\end{equation}
provides our estimate of the size of the SB system. The sum is limited to the $N$ largest firms. We take $N=1000$ but the results depend very weakly on the choice of $N$ as long as $S_{[N]}$ is in the range $[S_-,S_+]$ over which the fit is made, because $\hat S_{[k]}\simeq S_{[k]}$ within this range.

\begin{figure}[h]
   \centering
  \includegraphics[width=4in]{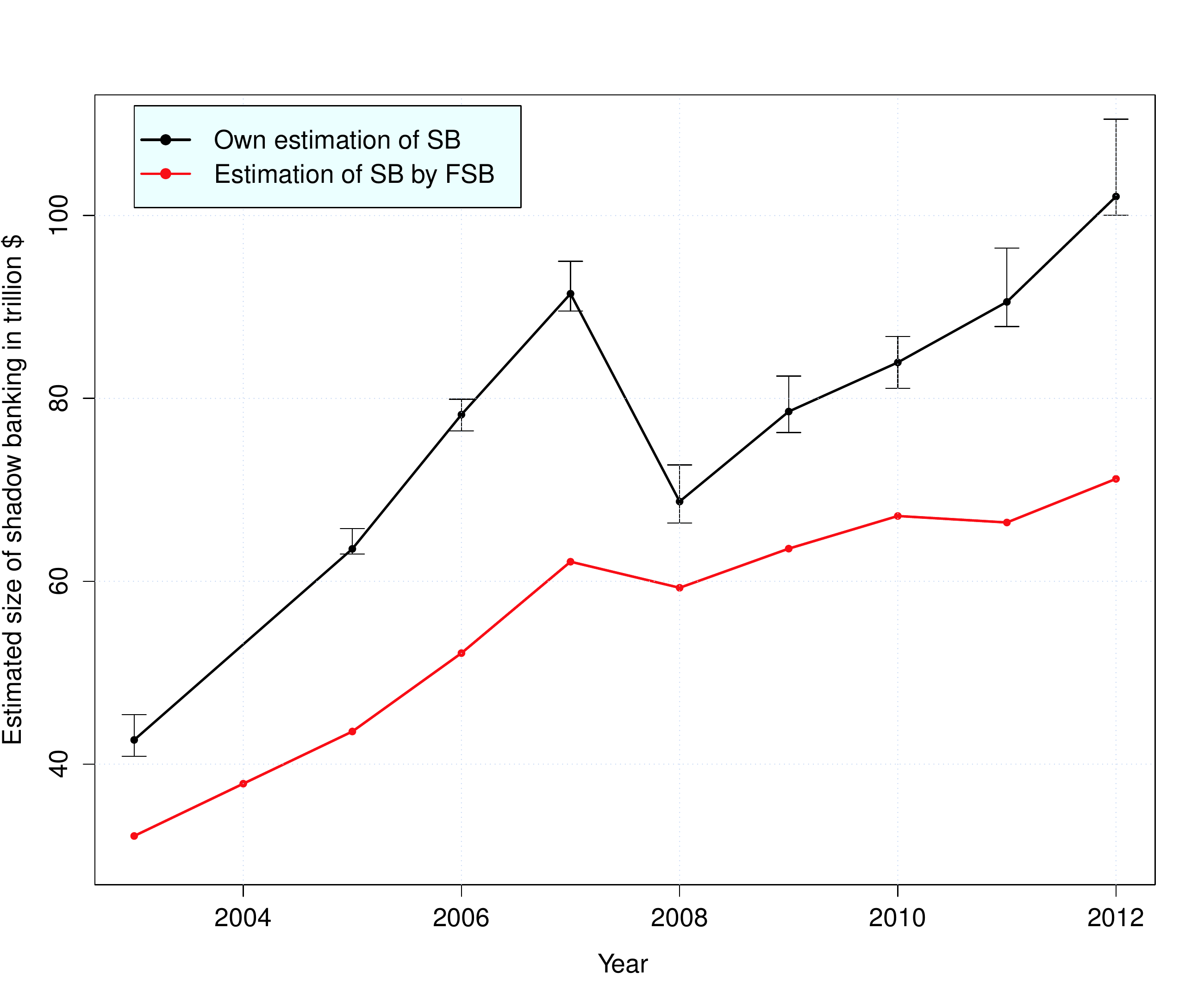}
   \caption{Comparison between our index of SB, $I_{SB}$, with the estimate of the size of SB made by FSB \cite{FSB2013b} for the period 2003-2012. The reported confidence bands for our estimate of SB are calculated on the basis of $\pm 2$ standard errors in the estimate of the coefficients of the power law distributions.}
   \label{fig:SBIndex_comparison}
\end{figure}

For comparison, Fig. \ref{fig:SBIndex_comparison} reports also the estimated size of the SB system by FSB  \cite{FSB2013b}. Both their estimate and $I_{SB}$ show a strong rise before the crisis in 2007, a drop in 2008 (much more severe for $I_{SB}$), and a growth after 2008, with $I_{SB}$ increasing at a faster pace, especially in 2011.

Ref. \cite{Gravelle2013} argues that an entity-based approach to SB, such as that of the FSB, "[...] may omit SB activities undertaken by banks that may contribute to systemic risk." Furthermore, as observed by Adrian {\em et al.} \cite{Adrianetal2013} "[...] the shadow banking system comprises many different entities and activities. In addition, the types of entities and activities which are of particular concern will change in the future, in response to new regulations". Along similar lines, Pozsar {\em et al} \cite{PozsarEtAl2013} conclude: "[...] the reform effort has done little to address the tendency of large institutional cash pools to form outside the banking system. Thus, we expect shadow banking to be a significant part of the financial system, although almost certainly in a different form, for the foreseeable future". 
These arguments suggest that the FSB estimate is likely to provide a lower bound to the size of the SB system. $I_{SB}$ may instead be considered as a \textit{theoretical upper bound}, as it measures the amount of assets that are missing from a hypothetical economy in which PRG holds across all scales of asset sizes.

A few comments are in order about $I_{SB}$:
\begin{itemize}
  \item $I_{SB}$ is a genuine systemic indicator, as it depends on a collective property of the economy. It is hard to manipulate and simple to compute, as it requires only data publicly available.
   \item $I_{SB}$ does not rely on a detailed list of entities and/or activities which contribute to the SB system; it is therefore robust to change in regulation and fiscal policy.
  \item $I_{SB}$ implicitly attributes SB activities to the largest financial firms which populate the top tail of assets distribution. It is well documented that the main financial firms originated most of the SB activities before the crisis \cite{Fein2013}. Yet, $I_{SB}$ also crucially depends on the exponent $\gamma$, whose estimate depends on the shape of the distribution in the intermediate range. In particular, \textit{ceteris paribus}, $I_{SB}$ is expected to increase if the exponent $\gamma$ decreases and {\em vice-versa}. 
\end{itemize}

A comparison between Table \ref{tab:estimatedParetoCoeff} and Fig. \ref{fig:SBIndex_comparison} shows how $I_{SB}$ is (anti)correlated with $\hat\gamma$ and $\hat{\gamma}_{\rm fin}$: when the assets distribution gets broader (i.e. $\gamma$ and $\gamma_{\rm fin}$ decrease), $I_{SB}$ increases and {\em vice-versa}. After the 2007-08 crisis, the correlation of $I_{SB}$ with $\hat{\gamma}_{\rm fin}$ is much stronger than with $\hat\gamma$. This is a further indication that the behavior of financial firms is at the core of the dynamics of $I_{SB}$.

\subsubsection*{A Simple Proportional Random Growth Model with Shadow Banking}
\label{sec:model}

Malevergne {\em et al.} \cite{Sornette2013} discuss a simple PRG model which displays a Paretian distribution. In their notation \cite{Sornette2013}, firms grow according to a log-normal stochastic process with drift $\mu$ and variance $\sigma$, they disappear from the market according to a Poisson point process at rate $h$, and new firms enter the market at rate $\nu$. In the following we neglect the possibility of an exogenous growth of economy (i.e. we take $c_0=d=0$ in the notation of \cite{Sornette2013}), and we assume that all new firms have initial size equal to one. Malevergne {\em et al.} \cite{Sornette2013} show that the top tail of the equilibrium size distribution has a power law shape with a Pareto exponent given by Eq. (7) in Ref. \cite{Sornette2013}:
\begin{equation}
\gamma = \frac{1}{2} \left[ \left(1 -2 \frac{\mu}{\sigma^2}\right) + \sqrt{\left( 1 -2 \frac{\mu}{\sigma^2} \right)^2 + 8 \frac{h}{\sigma^2}} \, \right] .
\label{eq:index}
\end{equation}
We introduce a modification of the original model, in order to reproduce the cut-off displayed by the Forbes data, i.e. in the presence of SB. According to a Poisson point process at rate $\lambda$, the largest firm $i^*$ in the economy (with size $S_{i^*}=\max_i S_i$) moves a fraction $\epsilon$ of its assets outside the regular banking system to the SB system, reducing its \textit{observed} size to $(1-\epsilon) S_{i^*}$. The parameter $\lambda$ is therefore a proxy for the \textit{intensity} of the activity feeding the SB system.

\begin{figure}[htbp]
   \centering
   \includegraphics[width=4in]{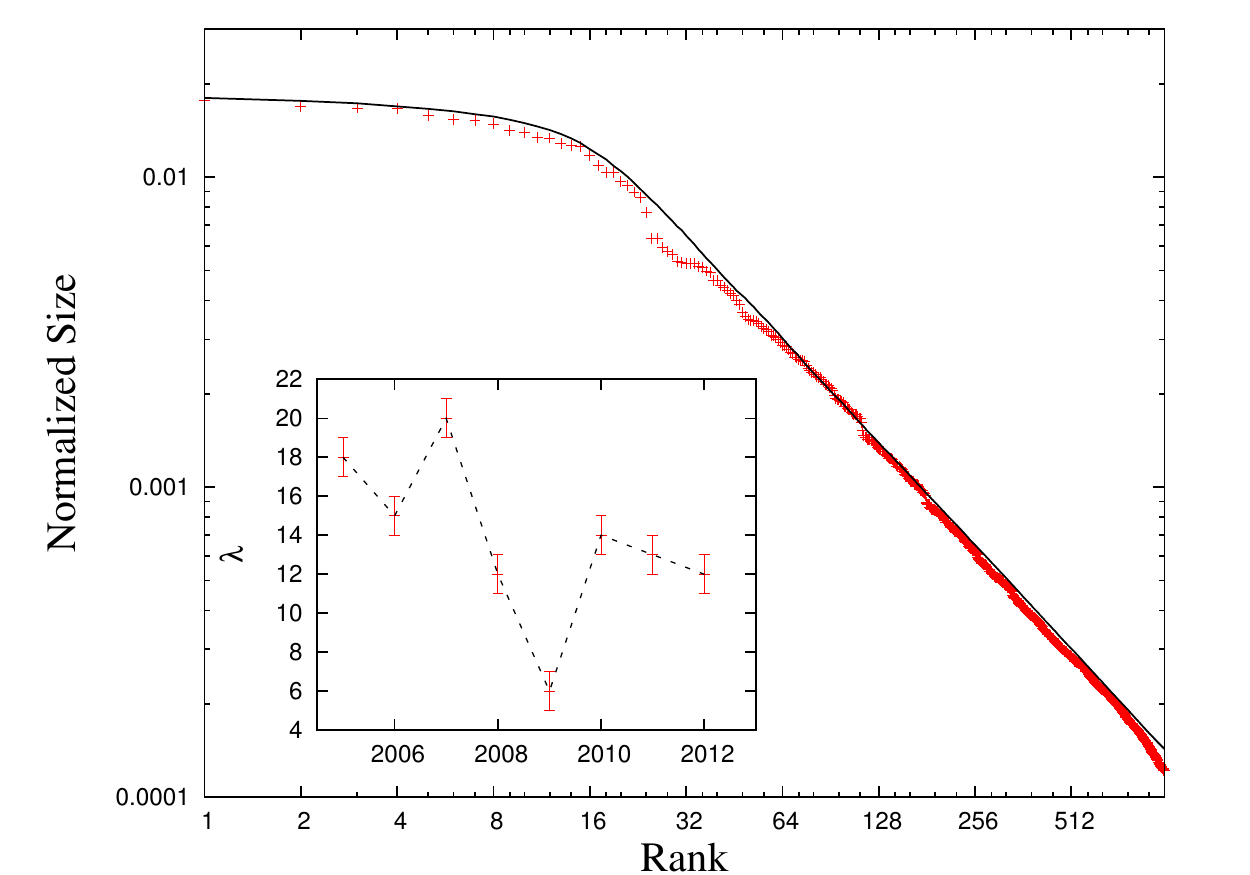}
   \caption{Comparison between the observed (cross) and the simulated (bold line) distributions for 2012. 
   Inset: the estimate of $\lambda$ in the period 2005-2012.}
   \label{fig:simulationVsObserved}
\label{fig:model}
\end{figure}

Table \ref{tab:lambda} reports a calibration of the parameters of the modified PRG model for the period 2005-2012 based on the FG2000 list of firms (see caption of the table). We set $\epsilon=0.1$ and we located the value of $\lambda$ that yields the best match between the simulated and the observed firm size distributions (see caption of Table \ref{tab:lambda}). Figure \ref{fig:simulationVsObserved} shows the quality of our calibration of the model for 2012.
The same fitting procedure was performed for different values of $\epsilon$; for $\epsilon \in \left(0,0.1\right)$ the ``flux'' $\epsilon\lambda$ of capital flow into the SB system results independent of $\epsilon$. This is reasonable, because when $\epsilon$ is very small and $\lambda$ very large, wealth is repeatedly drawn into the SB system from the same firm (the largest one). 

According to the estimate of $\lambda$ reported in Table \ref{tab:lambda}, the \textit{intensity} of SB activity peaked in 2007 before the financial crisis, when the {\em originate-to-distribute} activities implemented by asset-backed securities and other credit derivatives probably reached their zenith \cite{PozsarEtAl2013}. In 2008 and 2009 the SB activity showed a dramatic fall in agreement with the sharp decline in all economic activities and the supposed breakdown of the SB system; but from 2010 to 2012 we observe a renewed increase, even though not at pre-crisis rates. This dynamic is fully consistent with the evolution of the size of SB system reported in Fig. \ref{fig:SBIndex_comparison}, being $\lambda$ a proxy for the \textit{intensity} of the activity feeding the SB system. Considering that the largest firm is of the order of \$ 3 trillion in 2013, our result $\epsilon\lambda\approx 1.2$ suggests a flow of capital into the SB system that is currently progressing at approximately \$ 3.5 trillions a year.

\section*{Conclusions and Outlook}
\label{sec:conclusionsOutlook}
  
This paper takes a non-standard approach to studying the properties of an economy. 
Based on solid evidence in the literature \cite{Gabaix2009}, we consider the Pareto distribution for asset sizes as an empirical law of an economy. The observation of power law distributions in economics is a remarkably solid piece of empirical evidence, dating back to the work of Pareto \cite{Pareto}. This empirical law arises from a generic mechanism --  proportional random growth -- that is expected to work in particular for financial firms. The actual distribution of firm sizes, at the global scale, closely follows this empirical law in the middle range, but deviates markedly from it in the upper tail, which is populated entirely by financial firms. 
We invoke SB as the element that would reconcile observations with the expected law. This allows us to derive an index that identifies the size of SB with the missing mass in the top tail of the asset size distribution. This approach resembles the one leading astrophysicists to invoke {\em dark matter} and {\em dark energy} in order to reconcile empirical observations with the law of gravitation (current estimates suggest that dark matter and dark energy account for approximately 95\% of the total mass in the universe). Likewise, the observation of a truncated power law in the distribution of asset sizes, points to the existence of {\em dark assets} that account for the missing mass in the top tail of the distribution.

Our estimate of the SB size is silent about the precise nature of SB activities and entities, as well as about the mechanisms that generate the observed departure from the theoretical power law behavior. The missing mass from the top tail of the distribution does not necessarily correspond to hidden assets. It may rather refer to assets being redistributed within the system. The creation of Special Investment Vehicles in the securitization process is one example of a mechanism that transfers assets from large banks in the top tail to the bulk of the distribution.

The index is based on a simple and robust statistical feature, that is expected to characterize the collective behavior of an economy.   
Andrew Haldane \cite{Haldane2012} recently argued that monitoring and regulation based on a detailed classification of financial activities is unlikely to keep pace with the rate of innovations in the financial industry.
The increase in complexity of financial markets should rather be tamed by measures based on simple metrics. The index of SB proposed in this paper is a contribution in this direction.

Our study also raises a number of issues. We conjecture that China's financial sector may account, at least in part, for the disparity between our index of SB and the estimate reported by the FSB. Indeed, S\&P estimated that the outstanding Chinese SB credit totaled \$3.8 trillion by the end of 2012, which is 34\% of on-balance-sheet loans and 44\% of China's GDP \cite{news_on_china}. This estimate is 5.6 times larger than the FSB estimate of China's SB in 2011. The latest report of FSB \cite{FSB2013b} at the time of writing acknowledges the rapid growth in asset size of "Other Financial Intermediaries" in China (by 42\%) as well as in emerging market jurisdictions.
Chinese firms are rapidly growing in an environment that, in turn, is also changing very quickly, with features not always transparent or well understood (see, e.g., \cite{news_on_china}). 

On the theoretical side, inspired by Ref. \cite{Mehrlingetal2013}, in Section \ref{sec:model} we discuss a PRG model reproducing the observed behavior of the largest financial firms based on an {\em originate-to-distribute} activity, by which firms at the top of the distribution of firm sizes shift part of their assets off-balance-sheet, e.g. with the creation of Special Purpose Vehicles. Within this framework, we estimate the intensity of SB activity in 2005-2012, which largely agrees with the observed behavior of the SB system. In this respect, one promising direction of research which may provide clues to the r\^ole of finance in our global economy is to study the relationship between the fast growth of financial firms relative to non-financial firms and the proliferation of financial instruments, as in Ref. \cite{Marsili2013}.

\section*{Materials and methods}

Here we provide some discussion of the data used, which is publicly available at http:$//$www.forbes.com$/$global2000$/$list$/$ (FG2000) and at http:$//$money.cnn.com$/$magazines$/$fortune$/$fortune500$/$2013$/$full$\_$list$/$ (Fortune 500).
The FG2000 list refers to the previous year. Thus the 2013 FG2000 list collects firms according to their characteristics in 2012. In the present paper the financial sector includes all the firms that in the FG2000 list belong to the following industries: Banking, Diversified Financials, Insurance, Consumer Financial Services, Diversified Insurance, Insurance Brokers, Investment Services, Major Banks, Regional Banks, Rental \& Leasing, Life \& Health Insurance, Thrifts \& Mortgage Finance, Property \& Casualty Insurance. Their number ranges from 501 in the 2013 list to 597 in the 2008 list.

2004 is the first year we could find for the FG2000 list and 1995 is the first year when Forbes 500 began to include also financial firms. Even though it is classified as non-financial, General Electric is one of the largest issuers of commercial paper in the US and over 80\% of its assets are in the financial sector. Other non-financial firms that appear at the top of the list belong to the car industry, telecommunications and energy. While Ford and General Motors dropped by more than 100 places in the list by asset size, Volkswagen climbed from 108th to 74th. Vodafone also declined from 56th in 2004 to 125th in the 2013 list (by asset size). Oil and gas companies (BP, Exxon, Royal Dutch Shell), on the other hand, kept a remarkably stable position around the 85th place.

Assets of the non-financial firms in FG2000 totaled approximately \$20 trillion both in the 2004 and 2013 lists, whereas the total assets of financial firms in FG2000 list increased steadily from \$48 trillion in 2004 to \$138 trillion in 2013, twice the world's GDP. 
This trend is called {\em financial deepening} in Ref. \cite{Haldane2012b}, to which we refer for a discussion on the systemic implication of the growth in the size of banks.

All computations were made in R, see Ref. \cite{R2013}. All datasets and codes are available upon request. In particular, the nonparametric estimate reported in the left panel of Figure \ref{fig:retuns_to_scale_fin} (a Nadaraya-Watson kernel regression) was made with the R package \cite{sm2013}. The estimate of the stochastic kernel (i.e. of the conditional probability distribution) in the right panel of Figure \ref{fig:retuns_to_scale_fin} was obtained using the \textit{adaptive kernel estimation} discussed in Ref. \cite{FiaschiRomanelli2009}.

\section*{Acknowledgments}

We thank Kartik Anand, Joseph A. Langsam, Aldo Nassig, Tomohiro Ota, Zoltan Pozsar, Didier Sornette and Istvan P. Szekely for useful comments and discussions. We gratefully acknowledge their feedback, but remark that they are in no way responsible for the message expressed in our paper. We are obliged to Ann Eggington for checking the English of the paper.
I.K. is grateful to the Abdus Salam ICTP, Trieste, for the hospitality extended to him at the initial phase of this work. 

\bibliography{biblio}

\begin{thebibliography}{10}
\providecommand{\url}[1]{\texttt{#1}}
\providecommand{\urlprefix}{URL }
\expandafter\ifx\csname urlstyle\endcsname\relax
  \providecommand{\doi}[1]{doi:\discretionary{}{}{}#1}\else
  \providecommand{\doi}{doi:\discretionary{}{}{}\begingroup
  \urlstyle{rm}\Url}\fi
\providecommand{\bibAnnoteFile}[1]{%
  \IfFileExists{#1}{\begin{quotation}\noindent\textsc{Key:} #1\\
  \textsc{Annotation:}\ \input{#1}\end{quotation}}{}}
\providecommand{\bibAnnote}[2]{%
  \begin{quotation}\noindent\textsc{Key:} #1\\
  \textsc{Annotation:}\ #2\end{quotation}}
\providecommand{\eprint}[2][]{\url{#2}}

\bibitem{Axtell2001}
Axtell R (2001) Zipf distribution of u.s. firm sizes.
\newblock Science 293: 1818-1820.
\bibAnnoteFile{Axtell2001}

\bibitem{Fujiwara2004}
Fujiwara Y (2004) Zipf law in firms bankruptcy.
\newblock Phys A 337: 219-230.
\bibAnnoteFile{Fujiwara2004}

\bibitem{Gabaix2009}
Gabaix X (2009) Power laws in economics and finance.
\newblock Annual Review of Economics 1: 255-293.
\bibAnnoteFile{Gabaix2009}

\bibitem{BouchaudMezard2000}
Bouchaud J, Mezard M (2000) Wealth condensation in a simple model of economy.
\newblock Physica A 282: 536-545.
\bibAnnoteFile{BouchaudMezard2000}

\bibitem{Sornette2013}
Malevergne Y, Saichev A, Sornette D (2013) Zipf's law and maximum sustainable
  growth.
\newblock J Econ Dyn and Control 37: 1195-1212.
\bibAnnoteFile{Sornette2013}

\bibitem{FSB2012}
Board FS (2012).
\newblock Global shadow banking monitoring report.
\newblock \urlprefix\url{http://www.financialstabilityboard.org/}.
\bibAnnoteFile{FSB2012}

\bibitem{NewsEU2013}
Commission E (2013).
\bibAnnoteFile{NewsEU2013}

\bibitem{FSB2013}
Board FS (2013).
\newblock Fsb chair's letter to g20 leaders on progress of financial reforms,
  sept. $5^{\rm th}$ 2013.
\bibAnnoteFile{FSB2013}

\bibitem{WheelockWilson2012}
Wheelock DC, Wilson PW (2012) Do large banks have lower costs? new estimates of
  returns to scale for u.s. banks.
\newblock Journal of Money, Credit and Banking 44(1): 171-199.
\bibAnnoteFile{WheelockWilson2012}

\bibitem{RestrepoEtAl2013}
Restrepo DA, Kumbhakar SC, Sun K (2013) Are us commercial banks too big?
\newblock Universidad EAFIT: Documentos de trabajo 13-14.
\bibAnnoteFile{RestrepoEtAl2013}

\bibitem{Cristelli2012}
Cristelli M, Batty M, Pietronero L (2012) There is more than a power law in
  zipf.
\newblock Sci Rep 2: 812.
\bibAnnoteFile{Cristelli2012}

\bibitem{McCulley2007}
McCulley P.
\newblock Pimco global central bank focus.
\bibAnnoteFile{McCulley2007}

\bibitem{Fein2013}
Fein ML (2013).
\newblock The shadow banking charade.
\newblock \urlprefix\url{http://ssrn.com/abstract=2218812}.
\bibAnnoteFile{Fein2013}

\bibitem{PozsarEtAl2013}
Pozsar Z, Adrian T, Ashcraft A, Boesky H (2013) Shadow banking.
\newblock forthcoming in Federal Reserve Bank of New York: Economic Policy
  Review .
\bibAnnoteFile{PozsarEtAl2013}

\bibitem{Pozsar2008}
Pozsar Z (2008) The rise and fall of the shadow banking system.
\newblock Moody's Economycom .
\bibAnnoteFile{Pozsar2008}

\bibitem{Mehrlingetal2013}
Mehrling P, Pozsar Z, Sweeney J, Neilson DH (2013).
\newblock Bagehot was a shadow banker: Shadow banking, central banking, and the
  future of global finance.
\bibAnnoteFile{Mehrlingetal2013}

\bibitem{Gravelle2013}
Gravelle~T TG, Lavoie S (2013).
\newblock {\em Monitoring and Assessing Risks in Canada's Shadow Banking
  Sector}, bank of canada, financial system review, june 2013.
\bibAnnoteFile{Gravelle2013}

\bibitem{Deloitte}
Kocjan J, Ogilvie D, Schneider A, Srinivas V (2012).
\newblock The deloitte shadow banking index: Shedding light on banking shadows.
\bibAnnoteFile{Deloitte}

\bibitem{FSB2013b}
Board FS (2013).
\newblock Global shadow banking monitoring report (2013), nov. $14^{\rm th}$
  2013.
\bibAnnoteFile{FSB2013b}

\bibitem{Adrianetal2013}
Adrian T, Covitz D, Liang N (2013).
\newblock Financial stability monitoring, finance and economics discussion
  series divisions of research \& statistics and monetary affairs, federal
  reserve board, washington, d.c.
\bibAnnoteFile{Adrianetal2013}

\bibitem{Pareto}
Pareto V (1896).
\newblock Cours d'\'economie politique profess\'e a l'universit\'e de lausanne.
\bibAnnoteFile{Pareto}

\bibitem{Haldane2012}
G HA, Madouros V (2012).
\newblock Bis central bankers' speech at federal reserve bank of kansas city,
  366th economic policy symposium, the changing policy landscape, jackson hole,
  wyoming, 31 august 2012.
\bibAnnoteFile{Haldane2012}

\bibitem{news_on_china}
Yao K (2013) China banks significantly exposed to shadow financing: Fitch.
\newblock Reuters, April 10, 2013 .
\bibAnnoteFile{news_on_china}

\bibitem{Marsili2013}
Marsili M (2013) Complexity and financial stability in a large random economy.
\newblock forthcoming in Quantitative Finance (published online 25 Jun 2013) .
\bibAnnoteFile{Marsili2013}

\bibitem{Haldane2012b}
G HA (2012).
\newblock The 2012 beesley lectures, at the institute of directors, london, 25
  october 2012.
\bibAnnoteFile{Haldane2012b}

\bibitem{R2013}
Team RC (2013) R: A language and environment for statistical computing.
\newblock R Foundation for Statistical Computing .
\bibAnnoteFile{R2013}

\bibitem{sm2013}
Bowman AW, Azzalini A (2013).
\newblock \textit{R} package \texttt{sm}: {\em nonparametric smoothing methods
  (version 2.2-5)}.
\bibAnnoteFile{sm2013}

\bibitem{FiaschiRomanelli2009}
Fiaschi D, Romanelli M (2009) Nonlinear dynamics in welfare and the evolution
  of world inequality.
\newblock Temi di discussione (Economic working papers) 724, Bank of Italy,
  Economic Research and International Relations Area .
\bibAnnoteFile{FiaschiRomanelli2009}

\end{thebibliography}


\newpage
\section*{Tables}

\begin{table}[ht]
\centering
\begin{tabular}{crrcc}
  \hline
   \hline
List FG2000 & $S_-$ & $S_+$ & $\hat{\gamma}$ & $\hat{\gamma}_{\rm fin}$ \\ 
  \hline
 2004 & 14.88 & 665.14 & 0.926   & 0.710\\ 
 		 & 			  &				& (0.0012) &  (0.0019) \\
2006 & 11.02 & 897.85 & 0.889 & 0.678\\ 
 		 & 			  &				& (0.0005) &  (0.0013) \\
2007 & 12.18 & 992.27 & 0.871 & 0.645 \\ 
 		 & 			  &				& (0.0005) &  (0.0012) \\
2008 & 12.18 & 1096.63 & 0.864 & 0.655\\ 
 		 & 			  &				& (0.0006) &  (0.0016) \\
2009 & 14.88 & 1339.43 & 0.899 & 0.672\\ 
 		 & 			  &				& (0.0008) &  (0.0012) \\
2010 & 14.88 & 1339.43 & 0.891 & 0.674\\ 
 		 & 			  &				& (0.0008) &  (0.0011) \\
2011 & 18.17 & 1339.43 & 0.899 & 0.669\\ 
 		 & 			  &				& (0.0006) &  (0.0013) \\
2012 & 24.53 & 1635.98 & 0.905 & 0.648\\ 
 		 & 			  &				& (0.0009) &  (0.0012) \\
2013 & 24.53 & 1998.20 & 0.897 & 0.627\\ 
 		 & 			  &				& (0.0008) &  (0.0009) \\
   \hline
    \hline
\end{tabular}
\caption{The range of assets (in billion \$) $[S_-,S_+]$ where the power law behavior is estimated (for the whole sample), and the estimated Pareto exponents $\hat{\gamma}$ both for the whole sample and limited to the financial firms in the FG2000 lists from 2004 to 2013 (data for 2005 are not available). Standard errors of the estimated Pareto exponents are reported in brackets.}
\label{tab:estimatedParetoCoeff}
\end{table}

\begin{table}
\begin{center}
    \begin{tabular}{rrrrrrr}
    \hline
     \hline
Year & $\mu$ & $\sigma$ & $\gamma$ & $h$ & $\epsilon$ & $\lambda$ \\
\hline
2005  & 0.12 & 0.20 & 0.89 & 0.10 & 0.1 & 18 \\ 
2006  & 0.10 & 0.24 & 0.87 & 0.09 & 0.1 & 15 \\ 
2007  & 0.15 & 0.22 & 0.86 & 0.13 & 0.1 & 20 \\ 
2008  & 0.11 & 0.28 & 0.90 & 0.10 & 0.1 & 12 \\ 
2009  & 0.04 & 0.21 & 0.89 & 0.04 & 0.1 & 6 \\ 
2010  & 0.11 & 0.23 & 0.90 & 0.10 & 0.1 & 14 \\ 
2011  & 0.10 & 0.17 & 0.91 & 0.09 & 0.1 & 13 \\ 
2012  & 0.09 & 0.17 & 0.90 & 0.08 & 0.1 & 12  \\ 
    \hline
    \hline
    \end{tabular}
    \caption{Estimates of the parameters of the modified PRG model of the SB system for the period 2005-2012 based on the FG2000 list of firms.
    $\mu$ and $\sigma$ are calculated by yearly variations of the firms' asset size between consecutive years, except for 2005 we use data of 2003, instead of 2004 which is missing. Using these values, $h$ is computed from the estimate of the Pareto index $\gamma$, inverting Eq. (\ref{eq:index}).
    The reported value of $\lambda$ is the one that minimizes the distance between the observed and the simulated firm size distributions. Specifically, {\em i)} we compute $Z_k =\langle \log \left(  S_k/S_k^0 \right)\rangle$ with $S_k$ being the $k$-th largest firm in the simulation, $S_k^0$ the $k$-th largest firm in the FG2000 list and $\langle \, \cdot \, \rangle$ is the average over 100 simulations. {\em ii)} We find $\lambda$ that minimizes the mean square deviation $\sum_k (Z_k-\bar Z)^2/N$, with $\bar Z=\sum_kZ_k/N$. 
    }
    \label{tab:lambda}
\end{center}
\end{table}


\end{document}